\documentclass[12pt]{article}

\usepackage{graphicx}
\usepackage{amsmath}
\begin{document}

\begin{center}
{\large\bf Efficient Simulation of Fluid Flow and Transport in Heterogeneous Media Using Graphics 
Processing Units (GPUs)}

\bigskip
Hassan Dashtian and Muhammad Sahimi$^\dagger$

{\it Mork Family Department of Chemical Engineering and Materials Science, University of Southern 
California, Los Angeles, California 90089-1211}

\end{center}

Networks of interconnected resistors, springs and beams, or pores are standard models of studying scalar 
and vector transport processes in heterogeneous materials and media, such as fluid flow in porous media, 
and conduction, deformations, and electric and dielectric breakdown in heterogeneous solids. 
The computation time and required memory are two limiting factors that hinder the scalability of the 
computations to very large sizes. We present a dual approach, based on the use of a combination of the
central processing units (CPUs) and graphics processing units (GPUs), to simulation of flow, transport,
and similar problems using the network models. A mixed-precision algorithm, together with the 
conjugate-gradient method is implemented on a single GPU solver. The efficiency of the method is tested 
with a variety of cases, including pore- and random-resistor network models in which the conductances are
long-range correlated, and also contain percolation disorder. Both isotropic and anisotropic networks are
considered. To put the method to a stringent test, the long-range correlations are generated by a 
fractional Brownian motion (FBM), which we generate by a message-passing interface method. For all the 
cases studied an overall speed-up factor of about one order of magnitude or better is obtained, which 
increases with the size of the network. Even the critical slow-down in networks near the percolation 
threshold does not decrease the speed-up significantly. We also obtain approximate but accurate bounds for
the permeability anisotropy $K_x/K_y$ for stratified porous media.

\bigskip

\bigskip

\noindent\rule{4truecm}{0.01in}

\noindent $^\dagger$E-mail: moe$@$usc.edu

\newpage

\begin{center}
{\bf I. INTRODUCTION}
\end{center}

Fluid flow and transport in heterogeneous media is an important problem [1], in view of its relevance to 
a wide variety of phenomena in natural and industrial processes. A partial list of such phenomena include
flow through porous media [2,3], conduction and hopping transport in composite solids [4,5], mechanical 
properties of disordered materials [1,5], fracture of heterogeneous solids [6-8], and many more. For over
four decades the standard model of a heterogeneous medium has been a random resistor or pore network
(RRN and PN, respectively) in the case of flow and scalar transport, and a network of springs or beams 
for studying such vector transport processes as deformation and fracture propagation. Often, disorder is 
introduced in the model by a percolation process [9,10] by which a fraction of the bonds of the network -
pores, resistors, or springs - do not allow fluid flow, or scalar or vector transport to occur, while 
the rest of the bonds represent the fluid flow or transport paths. Even with such a relatively simple 
form of disorder a wide variety of interesting phenomena occur, a detailed understanding of which 
requires simulating very large networks and, thus, solving very large set of flow and transport 
equations. 

Numerous approaches, ranging from continuum to networks models, integrate [11] the complexity of 
heterogeneity of porous media with fluid flow. For example, the PN models discretize the void space of a 
porous structure into a network consisting of pore bodies (nodes or sites) connected via pore throats
(bonds) - hereafter referred to as, respectively, pores and throats - and have been widely used to 
investigate a wide variety of phenomena related to multiphase flow [12-16]. The applications includes 
drying of porous media [17-19], reactive transport [20-22], dissolution [23], unstable miscible 
displacements and fingering phenomenon [24], grain boundary wetting [25] and gas flow in shaly formations
[26,27]. The computational limitations of the PN modeling restrict, however, its application to 
relatively small networks. 

Thus, attention has also been focused on developing efficient methods for simulating very large PNs or 
RRNs and solving the associated large sets of flow or transport equations. The efforts has led to the 
development of the transfer-matrix method [28-30], and coarsening the disordered PNs or RRNs based on 
wavelet transformations and then solving a significantly-reduced number of equations [31]. With the 
exception of the wavelet coarsening method, which is efficient both near [33,34] and away from the 
percolation threshold [31,32], all the aforementioned methods are efficient only if the PN or RRN is near
the percolation threshold. Aghaei and Piri [35] developed a computational strategy for simulating 
multiphase flows in porous media that is capable of simulating large PNs. Their approach is, however, a 
massively parallel scheme.  

In practice, many heterogeneous media do not contain percolation-type disorder or, if they do, their
disorder is not random but highly correlated. Moreover, with the advent of sophisticated experimental 
techniques, such as x-ray computed tomography [36-38], it is now possible to obtain detailed data for the
morphology of heterogeneous media. Taking proper account of the correlations and utilizing the detailed 
data in the model entails employing a high-resolution PN or RRN with several million nodes, a still 
daunting task. Moreover, the problem is much more difficult when one must solve unsteady-state problems 
over a period of time, which entails solving a very large number of equations for thousands of time 
steps. 

In this paper we present an efficient approach to PN or RRN modeling, which is based on graphic 
processing units (GPUs) that have opened up a new approach for high performance computing, particularly 
for the researchers who do not have access to massively-parallel machines with a large number of nodes, 
or to vector supercomputers. The GPUs were originally designed to analyze three-dimensional graphic 
images. To achieve extremely high performance with geometric data, the GPUs have been designed with 
simple and tiny processors. More modules are used for data processing, but not for the data cache, nor 
for the flow control. Hence, by design the GPUs are very different from typical computations with 
central-processing units (CPU), and are ideal for highly intensive and parallelized calculations, if the 
computational algorithm can take advantage of their special features. In facilitating nongraphical 
calculations on its graphic units, Nvidia Corporation released Compute Unified Device Architecture (CUDA)
[39]. Several groups have tested the performance of CUDA parallel computing in various calculations 
[40-46]. 

\begin{center}
{\bf II. PORE AND RANDOM-RESISTOR NETWORKS}
\end{center}

The first step in the PN/RRN modeling is to generate the networks. It was suggested [3,47] a long time
ago that many natural porous media exhibit long-range correlations. Such correlations, which influence 
the percolation [48] and flow and transport properties of porous media [37,49-52], follow the statistics 
of fractional Browning motion (FBM) [53], a nonstationary stochastic process. Thus, in addition to a 
random distribution of the pores' or resistors' conductances, we also use the FBM to generate correlated 
conductances, or throats' size, in the PN or the RRN. We use a square network in which each bond 
represents a throat with an effective radius $r$, or a resistor, selected from a statistical 
distribution. The bonds' length $\ell$ is assumed to be constant, although it poses no difficulty to 
select it from a statistical distribution as well. The bonds' radii (or conductances) are selected from 
a FBM. An efficient method for generating a FBM array is through its power spectrum $S(\mbox{\boldmath$
\omega$})$, which for 2D systems is given by
\begin{equation}
S(\mbox{\boldmath$\omega$})=\frac{a}{(\omega_x^2+\omega_y^2)^{H+1}}\;,
\end{equation}
where $a$ is a constant, while $\omega_x$ and $\omega_y$ are the Fourier components in the $x$ and $y$ 
directions. Here, $H$ is the Hurst exponent such that for $H>0.5$ ($H<0.5$) one has positive (negative) 
correlations, while $H=0.5$ represents the case in which the successive increments of the FBM are 
completely random. Due to stratification, many porous media are anisotropic. To intrtroduce the 
anisotropy, we follow Ansari-Rad {\it et al.} [54] and generalize the power spectrum to
\begin{equation}
S(\omega)=\frac{b}{(\beta_x\omega_x^2+\beta_y\omega_y^2)^{H+1}}\;.
\end{equation}
Here, $\beta_x$ and $\beta_y$ are the anisotropy parameters, and $b$ is a constant. To generate 
anisotropy induced by layering with the layers being parallel to the $x$ direction, we set 
$\beta_x/\beta_y> 1$. For large PNs that we utilize in the present study, however, generating a large FBM
array is computationally expensive, because as the size of the PN increases, the computation and required
memory increase exponentially, if sequential programming is used for generating the FBM. Thus, to make 
the running time manageable, we used message-passing interface (MPI) strategy and generated the FBM 
algorithm as a parallel scheme in the GPU solver (see below).

The most time-consuming part of setting up a PN/RRN is generation of a 2D FBM by a fast Fourier transform
(FFT) that computes the spectral density, Eq. (2), with a time complexity of ${\cal O}(N\log_2N)$. 
Another main issue associated with the sequential implementation of the FBM algorithm is that, for large 
network sizes $N$, huge memory is required (see Table I below). To address the problem, we parallelized
the program by using the message-passing interface (MPI) strategy on four processes numbered (0, 1, 2, 3);
see Fig. 1. Each processor executes a part of the FBM generation algorithm, and then sends the results 
to the root process, number 0 in Fig. 1. In Table I we report the performance of the MPI implementation 
of the generation of the 2D FBM fields, and the PN/RRN of various sizes, which indicate that the 
efficiency and speed-up both increase as the size increases. Let us emphasize that generation of large
FBM arrays by the MPI strategy is, to our knowledge, new. Given the wide applicability of the FBM, our
algorithm makes it possible to generate very long FBM array, a difficult problem.

\begin{center}
\begin{table}[h]
\caption{Efficiency and speed-up of the MPI implementation of FBM algorithm on four processes. The times
are in seconds}
\begin{tabular}{cccccc}
\hline
Array Size & Required Memory (GB) & Time (Sequential) & Time (MPI) & Speed-up & Efficiency \\ \hline
$2^{24} $ & 0.128    & 1.2 & 0.4 & 2.8 & 0.71 \\ \hline
$2^{28} $ & 2 & 8.0  & 2.7 & 2.9 & 0.73 \\ \hline
$2^{30} $ & 8 & 71.6 & 20.6 & 3.5 & 0.87 \\ \hline
\end{tabular}
\end{table}
\end{center}

We then used the generated PN/RRN as the input to the GPU-based solver. Solving for the pressure/voltage
distribution consists of the CG algorithm, which is matrix operation, well suited for the GPUs. Several 
PNs/RRN were generated with various values of the Hurst exponent $H$, the anisotropy parameter
$\beta_x/\beta_y$, and network sizes $N$. In Fig. 2 we present samples of the generated PNs/RRN in which
the conductances vary over two orders of magnitude. 

Assuming steady-state and slow (laminar) flow of an incompressible fluid, the volume flow rate in a
bond $ij$ that connects nodes $i$ and $j$ is given by, $q_{ij}=\pi r^4\Delta P_{ij}/(8\mu\ell)\equiv
g_{ij}\Delta P_{ij}$, where $\mu$ is the fluid's viscosity, $\Delta P_{ij}=P_j-P_i$ is the pressure drop 
between nodes $i$ and $j$, and $g_{ij}$ the conductance of bond $ij$. Writing a mass balance for every 
node $i$, one obtains, $\sum_{j\in i}q_{ij}=0$, where the sum is over all the nodes $j$ that are 
connected to node $i$. Thus, substituting for $q_{ij}$ and writing the mass balance for every interior 
node of the network results in a set of linear equation of the following form,
\begin{equation}
{\bf G}{\bf P}={\bf b}\;,
\end{equation}
where {\bf G} is the conductance matrix that depends only on the geometric properties of the network, 
{\bf P} is the nodal pressure vector (or voltage vector in the RRN) to be calculated, and {\bf b} is a 
vector related to the external boundary conditions. We impose a fixed pressure gradient on the network 
in one direction, and used periodic boundary conditions in the second direction. Once the pressure 
(voltage) distribution in the PN is calculated, the effective permeability $K$ (conductivity) of the 
network is computed based on the Darcy's law,
\begin{equation}
K=\frac{Q}{S}\frac{\mu L}{\Delta P}\;,
\end{equation}
where $Q$ is the total flow rate passing through the cross section $S$ of the PN, $L$ is the PN's length,
and $\Delta P$ is the macroscopic pressure drop imposed on the network. A similar procedure is used to
compute the effective conductivity of the RRN. The set of the equations must be solved by the 
conjugate-gradient (CG) method, which is the standard method in the computations with the CPU, but must 
be implemented in the GPU. The matrix-vector multiplication is the most dominant part in any
CG iteration and, thus, it is this part that is accelerated by the GPU.

To implement the solution of the pressure/voltage distribution in a PN/RRN on GPUs, one must use CUDA 
programming that makes it possible to implement the CG algorithm on a single GPU. CUDA programming,
as well as the memory allocation and performance from the perspective of our GPU solver kernels, is by 
far the most time-consuming and resource-intensive computational step in the simulations. Thus, we first 
describe the implementation of the algorithm on GPUs, and present its technical details.

\begin{center}
{\bf III. CUDA and GPUs}
\end{center}

Memory allocation in, and performance of the implementation of the algorithm are by far the most 
time-consuming and resource-intensive computational steps. On a GPU, parallel tasks, called threads, are 
scheduled and executed simultaneously in groups referred to as warps. One warp contains 32 threads that
processed in parallel by one CUDA streaming multiprocessor (SM). The GPUs have many SMs that run in 
parallel to increase the parallelism. The threads are organized also into larger structures called 
blocks, which are themselves organized into grids [55]. Figure 3 shows an example of the GPU architecture
hierarchy with four SM.

The CUDA memory contains various segments with different scopes and properties. All the threads have 
access to the same global memory. The shared memory is visible to all the threads within a block with the
same lifetime as the block. Each thread has its own registers and local memory. The global memory is 
located in the GPU, which is the largest, but also the slowest. A programmer can allocate free global 
memory. In order to access the GPU memory, 32, 64, or 128 byte memory ``transactions'' are needed from 
the host to the device, which must have the same sizes, but because they are costly in terms of their 
computational performance, they must be aligned and minimized. In our implementation of the CG algorithm 
for solving the pressure/voltage distribution in a PN or RRN, we use 1D arrays to store the elements in 
a vector. The local memory is allocated by the compiler for large structures that do not fit into the 
register space, and are part of the global memory that can be used to avoid costly memory transfers. 

Despite local and global memories, shared memory is of the on-chip type and, hence, has much lower 
latency with higher bandwidth. The threads in a block can use shared memory to work together. For 
example, in the case that multiple threads in a block need to use the same data, shared memory is called 
upon to access the data from the global memory only once. The GPU that we use has a Kepler architecture 
[56], one of the most efficient high performance computing architecture in which the shared memory is 48 
KB, and is organized into 32 banks. In our CUDA kernel - functions that are executed on the GPU - we 
store temporal variables in the shared memory. In Fig. 4 we show the concept of memory in a typical GPU. 

\begin{center}
{\bf IV. SOLVING FOR THE PRESSURE/VOLTAGE DISTRIBUTION WITH A GPU}
\end{center}

We used the CG algorithm in both C++ and CUDA. In CUDA, the CPU is a host for GPU, with the latter called
a device. Programs that are run on the CPU (host) can manage the memory on both the CPU and GPU. The CPU
lunches the kernels, which and are then executed in a parallel scheme by the GPU threads. As far as the 
CG algorithm (as well as other iterative solvers) is concerned, the most computationally intensive kernel
is the matrix-vector multiplication (MVM) kernel [57]. Most of the execution time of the main loop of a 
solver is spent inside the MVM kernel.

One important issue related to the GPU and CPU is the floating point accuracy of computations. In order 
to ensure consistent computations across platforms and to exchange floating-point data, the IEEE-754 
defines basic and interchange formats to convert decimal floating point to 32 bit and 64 bit hexadecimal 
representations, along with their binary equivalents. The 32 and 64 bit basic binary floating point 
formats correspond to the C data types {\it float} and {\it double}. The GPU devices with high compute 
capability support both single and double precision (SP and DP, respectively) floating point computation.
On CUDA the compute capability is the ``feature set,'' for both the hardware and software, of the device.
The higher the compute capability, the more powerful the device. The GPU device that we use in this study
supports both the SP and DP.

There is, however, a significant gap between the SP and DP computations with the GPUs, as the SP 
calculations run much faster than the DP computations. The main bottleneck of the CG iterations is the 
ratio of the arithmetic operations and the data input/output steps, and the acceleration of the 
computations depends critically on this ratio. Thus, by using the SP data we reduced their ``traffic'' 
by a factor of two between the GPU processor and its device memory. Then, to further improve the accuracy
of the CG calculations, we used a mixed-precision method, implemented by an iterative refinement 
algorithm. The main idea is using two types of iterative loops to obtain the true solution of the 
equations. First, by using the SP iteration we approach rapidly a rough solution within the inner loop. 
Next, iterations with the DP are used to converge to the final solution within the allowed error of the 
outer loop. Since the iteration loop of the CG algorithm is responsible for most of the computation time,
we explain in detail the parallel implementation of this part on the GPU; see Fig. 5.

{\it First kernel}: A 1D block of size 256 is used. For each part, the size of the network is set 
proportional to the total number of rows and blocksize, such that there exists a thread corresponding to 
each row. 

{\it Second kernel}: This is the sparse matrix-vector multiplication part of the computation, for which a
thread is assigned to each row of matrix {\bf G}, which is responsible for calculating the elements of 
{\bf Z}, the vector that acts as temporary {\bf b} during the iteration process. Consecutive rows of 
{\bf G} have redundant access to {\bf P} through the calculations associated with the three main 
diagonals of {\bf G}. Thus, {\bf P} is cached in the shared memory for improving the access pattern. 
After the calculation of each entry of {\bf Z} in a thread, it is multiplied by its corresponding {\bf P}
element and the result is held in the shared memory. Then, by performing a reduction operation over each 
block, the sums, which arise when one multiplies one row of a matrix by the column of another matrix or a
vector, are calculated and stored in the global memory.

{\it Third Kernel}: For this part a thread is assigned for the calculation of each element of {\bf U}, 
{\bf R}, and {\bf Z} and, similar to the second kernel, partial result of dot products for {\bf R} and 
{\bf Z} are calculated and saved in the global memory for each block. Here, {\bf R} is the residual
vector for each iteration, while {\bf U} represents the temporary approximate solution after each 
iteration.

Merging the calculations of the vectors' entries into a single kernel leads to a more efficient 
performance by decreasing the number of times that the global memory must be accessed. We optimized the 
number of threads and blocks based on the computational capability of the GPU device that we used. 
The number of threads per block must be a multiple of the warp size, which is 32 in all the current 
hardwares. We used 256 threads per block and 64 blocks in a fixed 1D grid arrangement. A loop was placed 
inside the threads to calculate multiple outputs per thread, so that the first iteration through all the 
threads is responsible for the first $2^{14}$ row calculations; the next iteration for the next $2^{14}$ 
row calculations, and so on. 

After completion of the iteration loop, the results are saved in the shared memory. By performing a 
reduction operation in each block, the sum is calculated and saved in the global memory. As there are 
only $2^6$ blocks, there will be $2^6$ partial sums in the global memory, to be added together to 
calculate the final solution. By copying the sums' values to the main memory and performing the final sum
on the CPU, a more efficient performance is achieved. Thus, a multiprocess algorithm was designed in an 
attempt to use the global memory throughput more efficiently. Since most of the summation operations are 
performed in registers or the shared memory, there is little effect on the global memory bandwidth, and 
the efficacy of the mixed-precision design is not affected. 

Task latency - the elapsed time between initiation and completion of a task - and throughput - the rate 
at which the system can process tasks - are two fundamental measures of processor performance. Improving 
throughput or reducing latency results in a better speed-up. The CPUs perform better for latency-oriented
tasks, whereas the GPUs are designed for throughput-oriented computing systems. There are some specific 
measures that are used to decide whether a certain application (or task) is suitable for the GPU or the
CPU implementation. The {\it memory footprint} of a task, the amount of main memory that a program uses 
or references while running, is the primary measure. For applications that require large memory, the CPUs
can be equipped with more random access memory (RAM) to execute the application, whereas the GPUs are 
very limited in this regard. Other measures are used for evaluating parallel computations and their 
optimization. The fast memory access of the GPUs and their massively-parallel units are well suited for 
ordered data patterns. Vector operations perform better on the CPUs, whereas matrix operations are more
efficient on the GPUs. It is for such reasons that, as we explained in the main text of the paper,
we used a dual combination of CPU + GPU to carry out the computations.

We emphasize that we use a combination of CPU {\it and} GPU to carry out the PN/RRN simulation. Using 
such a combination of two distinct paradigms and programming models for solving a given problem is 
completely nontrivial. In our work we first generate the PN/RRN using four CPU processors, which are then
used in a GPU to solve for the pressure/voltage distribution in the PN/RRN. The percentage of run times 
spent on the CPU host vary with the size of the network, as well as the number of connected resistors or 
pores in the network. As the network's size increases, the run time on the CPU also increases due to the 
limited memory of the GPU. In the problems that we study the GPU code is memory-bounded and most of the 
time is spent on memory and communication between the nodes. As we show below, in this problem a 
considerable time is spent for the communications and memory transfer. What this means is that the 
efficiency will increase dramatically over what is presented in this paper, if the GPU memory also 
improves.

Our preliminary computations indicated that such a combination of the CPU and GPU yields the most 
efficient performance because of the following. The distribution of the data in a dual computing system, 
such as CPU and GPU, is an important factor that can restrict the performance. Although both CPUs and 
GPUs are capable of performing similar types of calculations, there are several important differences in 
their execution models and architectures. The number of cores on a CPU is limited, but they are large and
fast. GPUs, on the other hand, have hundreds of cores that are slower and, compared with the CPU cores, 
possess limited capability. It is expensive, in terms of the computer time, to transfer data from one 
processing unit from the CPU (GPU) to a unit in GPU (CPU), and vice versa. Generating a PN/RRN and 
solving for the pressure/voltage distribution, are, however, two independent problems that are solved 
sequentially. The fast memory access of the GPUs and their massively-parallel units are ideal for ordered
data patterns. Vector operations are carried out more efficiently on the CPUs, whereas matrix operations 
are executed faster on GPUs. Thus, it is efficient to use several CPUs to parallelize the PN/RRN 
generation, because it requires very large computer memory. 

The computations were carried out with various PN and RRN sizes in both CPU alone and in GPU, in order to
measure the speed-up of the computations. The computing system that we employed consisted of one Intel 
core i7-4710 CPU (2.5 GHz) with 16 GB of main memory, and the NVIDIA Geforce GTX 880 graphic card with 
1536 cores. All speed-up comparisons have been made with respect to runtime of sequential PN generation 
and simulation algorithm in C++. Applications were written in CUDA version 6.5 and C++ using Visual 
Studio 2012. 

\begin{center}
{\bf V. EFFICIENCY OF THE COMPUTATIONS}
\end{center}

Since one goal of this paper is demonstrating the efficiency of the computations with standard PN or RRN 
models, we put the algorithm to use in a most stringent environment. In the former case, the PN network 
contains long-range correlations between the bonds' permeabilities, as well as anisotropy, as described 
earlier. In the case of the RRN, we delete a fraction $q$ of the resistors and compute the voltage 
distribution. As $q$ approaches the percolation threshold of the network, $q=1/2$, there is usually
critical slow-down because the structure of the sample-spanning percolation cluster becomes increasingly 
tortuous. Thus, a key test of the method is whether the gained efficiency is lost as the percolation 
threshold is approached. 

For a 2D PN/RRN of size $N\times N$ the speed-up of the computations is defined as the ratio of the 
sequential execution time $t(1,N)$ and the corresponding parallel execution time $t({\cal P},N)$ with
${\cal P}$ processors,
\begin{equation}
S({\cal P})=\frac{t(1,N)}{t({\cal P},N)}\;.
\end{equation}
In the PN/RRN simulation, the parallel part consists of both the CPUs (for four processors) and the GPUs.
Thus, the parallel execution time is the sum of running time for {\bf both} generating the PN/RRN on 
the CPU and solving the governing equations on the GPU. Another way of measuring the performance of the
parallel implementation is the {\it efficiency} ${\cal E}({\cal P})$, defined by
\begin{equation}
{\cal E}({\cal P})=\frac{S({\cal P})}{{\cal P}}=\frac{t(1,N)}{{\cal P}t({\cal P},N)}\;.
\end{equation}
The efficiency is only used for the parallel implementation of PN/RRN generation.

\begin{center}
{\bf VI. RESULTS}
\end{center}

We carried outr extensive computations with a variety of PN and RRN in order to test the speed-up in the
computations. In what follows we present and discuss the results.

\begin{center}
{\bf A. Isotropic networks}
\end{center}

Figure 6 presents the effect of the Hurst exponent $H$ on the speed-up of computing the pressure/voltage 
distribution. The network is isotropic [$\beta_x=\beta_y$ in Eq. (2)], and each data point represents the
average of multiple realizations. The speed-up increases with increasing the size of the PN/RRN. For a 
network with $H=0.75$ and 3.6 million nodes, the speed-up is 12, more than an order of magnitude 
improvement in the computations. More importantly, the same speed-up is obtained with a FBM distribution 
with $H=0.35$, the case in which the correlations between the permeabilities of conductances are 
negative, which is typically the case in natural porous media [3]. Thus, the value of the Hurst exponent 
$H$ has only a minor effect, if at all, on the speed-up of the GPU-based solver, which is strong evidence
for the efficiency of the algorithm.

Next, we consider the case of the classical RRN model [58] in which a randomly-selected fraction 
$p=1-q$ of the bonds have a unit conductance, while the rest are insulating. The speed-ups are shown in 
Fig. 7. As the percolation threshold $q_c=p_c=1/2$ is approached, the speed-up decreases from about 9.6 
for $p=1$ to about 7.5 very close to $p_c$. Thus, even though, similar to any critical phenomenon, there 
is critical slow down in a RRN model as $p_c$ is approach, the speed-up obtained by using the GPU 
decreases by only about 20 percent, hence indicating the efficiency of the GPU-based computations.

\begin{center}
{\bf B. Anisotropic networks and permeability anisotropy}
\end{center}

As mentioned earlier, natural porous media, as well as many solid materials, are anisotropic. Thus, we 
studied the effect of the anisotropy parameter $\beta_x/\beta_y$ on the performance of CPU+GPU 
parallel computations. The results are shown in Fig. 8, where we present the speed-up for two values of
of $\beta_x/\beta_y$ and a Hurst exponent of $H=0.35$. The speed-up is essentially the same that of the
isotropic media, and also independent of the value of $\beta_x/\beta_y$. Thus, although in the 
traditional approaches the computations for anisotropic media take longer times, in the proposed 
algorithm both the Hurst exponent $H$ and the anisotropy parameter $\beta_x/\beta_y$ have a minor effect,
if any, on the speed-up gained by the dual CPU-GPU. 

Next, we consider the effect of percolation, i.e., attributing zero conductance to the bonds in PNs or
RRNs in which the conductances are correlated, with the correlations described by the FBM. The motivation
for considering such networks is that natural porous media, in addition to be mostly anisotropic, are 
highly heterogeneous and contain very low-permeability zones that contribute very little, if any, to 
fluid flow. Thus, their permeability or hydraulic conductance may be set to zero. Composite materials 
that consist of conducting and insulating phases behave the same. But, since the conductance field is 
correlated, one cannot remove bonds randomly, which would destroy the correlations. Instead, the bonds' 
conductances are sorted from the smallest to the largest, and a given fraction $q$ of the bonds with the 
smallest conductances are removed. The results are presented in Fig. 9. In all the case the network 
contained $3.6\times 10^6$ nodes. The speed-up for all the cases has a trend similar to Fig. 8. Note 
that as the percolation threshold $q_c=1/2$ is approached, the speed-up decreases. As pointed out 
earlier, this represents critical slow down as the percolation threshold is approached. However, even 
very close to $q_c$ the speed-up is nearly one order of magnitude.

An important problem in modeling of anisotropic porous media is the ratio of the permeabilities 
$K_x/K_y$, where $x$ represents the direction that is more or less parallel to the strata that make such 
porous media anisotropic. There are many empirical estimates of the ratio $K_x/K_y$ in the literature [3]
that vary between 1 and 7. Using our efficient solver, and generating the permeability distribution 
according to Eq. (2), we obtain approximate, but accurate estimates for the permeability anisotropy. The 
results are presented in Fig. 10. For $H=0.75$ the anisotropy varies between 0.6 and 1.6, whereas the 
corresponding numbers for $H=0.35$, i.e., the much more heterogeneous PNs, are 0.5 and about 2.5. 

\begin{center}
{\bf VII. DISCUSSION}
\end{center}

Several important points must be mentioned here. (i) Generally speaking, unlike CPU-based computations,
the denser the matrix {\bf G} is, the more efficient the GPU-based computations are. Thus, similar
calculations for 3D systems and with networks with higher connectivities should yield higher speed-ups. 
(ii) The calculations with the lattice models of vector transport, such as brittle fracture propagation 
and similar phenomena will be even more efficient than the scalar transport considered in this paper, 
because the corresponding matrix {\bf G} will be denser, even for 2D systems. (iii) Clearly, any of the
past methods of speeding-up the calculations will be more efficient if carried out with the GPUs. (iv) 
We utilized a single GPU-based solver. Use of several GPUs in parallel should result in very high 
speed-up factors, a matter that we are currently studying. 

\begin{center}
{\bf VIII. SUMMARY}
\end{center}

We proposed an algorithm for simulating the pore-network and random-resistor network models on GPUs. The 
single GPU-based solver improves the efficiency and running time of the calculations by at least one 
order of magnitude, when compared with the same calculations with CPU-based simulatorss. The speed-up is 
even larger for larger networks, and for calculations in which the matrix of the coefficients is denser. 
Thus, the GPU-based computations make simulation of very large PNs/RRNs possible, regardless of the 
correlations, anisotropy, or any other pertinent parameter.

\begin{center}
{\bf ACKNOWLEDGMENTS}
\end{center}

This work was supported as part of the Center for Geologic Storage of CO$_2$, an Energy Frontier Research
Center funded by the U.S. Department of Energy, Office of Science, Basic Energy Sciences, under Award 
number DE-SC0C12504.

\newpage

\newcounter{bean}
\begin{list}%
{[\arabic{bean}]}{\usecounter{bean}\setlength{\rightmargin}{\leftmargin}}

\item S. Torquato, {\it Random Heterogeneous Materials} (Springer, New York, 2002).

\item M.J. Blunt, {\it Multiphase Flow in Permeable Media} (Cambridge University Press, London, 2017).

\item M. Sahimi, {\it Flow and Transport in Porous Media and Fractured Rock}, 2nd ed. (Wiley-VCH, 
Weinheim, Germany, 2011).

\item B. I. Shklovskii and A.L. Efros, {\it Electronic Properties of Doped Semiconductors} (Springer, 
Berlin, 1984).

\item M. Sahimi, {\it Heterogeneous Materials I} (Springer, New York, 2003).

\item B. K. Chakrabarti and L. Benguigui, {\it Statistical Physics of Fracture and Breakdown in 
Disordered Systems} (Oxford University Press, London, 1997).

\item M. Sahimi, {\it Heterogeneous Materials II} (Springer, New York, 2003).

\item {\it Modelling Critical and Catastrophic Phenomena in Geoscience}, edited by P. Bhattacharyya and 
B. K. Chakrabarti (Springer, Berlin, 2006).

\item D. Stauffer and A. Aharony, {\it Introduction to Percolation Theory}, 2nd ed. (Taylor and Francis, 
London, 1994).

\item M. Sahimi, {\it Applications of Percolation Theory} (Taylor and Francis, London, 1994).

\item J. Chu, B. Engquist, M. Prodanovic, and R. Tsai, A multiscale method coupling network and continuum
models in porous media I: steady-state single phase flow, Multiscale Model. Simul. {\bf 10}, 515 (2012).

\item P.E. {\O}ren, S. Bakke, and O.J. Arntzen, Extending predictive capabilities to network models,
Soc. Pet. Eng. J. {\bf 3}, 324 (1998).

\item S. Ovaysi and M. Piri, Direct pore-level modeling of incompressible fluid flow in porous media,
J. Comput. Phys. {\bf 229}, 7456 (2010).

\item M. Piri and M.J. Blunt, Three-dimensional mixed-wet random pore-scale network modeling of two- and 
three-phase flow in porous media. I. Model description, {\it Phys. Rev. E} {\bf 71}, 026301 (2005). 

\item M. Piri and M.J. Blunt, Three-dimensional mixed-wet random pore-scale network modeling of two- and 
three-phase flow in porous media. II. Results, Phys. Rev. E {\bf 71}, 026302 (2005).

\item M.J. Blunt, M.D. Jackson, M. Piri, and P.H. Valvatne, Detailed physics, predictive capabilities and
macroscopic consequences for pore-network models of multiphase flow, Adv. Water Resour. {\bf 25}, 1069 
(2002).

\item M. Prat, Recent advances in pore-scale models for drying of porous media, Chem. Eng. J. {\bf 86}, 
153 (2002).

\item O. Borgman, P. Fantinel, W. Luhder, L. Goehring, and R. Holtzman, Impact of spatially correlated 
pore-scale heterogeneity on drying porous media, Water Resour. Res. {\bf 53}, 56455658 (2017).

\item A.G. Yiotis, D. Salin, and Y. C. Yortsos, Pore network modeling of drying processes in macroporous 
materials: Effects of gravity, mass boundary layer and pore microstructure, Transp. Porous Media 
{\bf 110}, 175 (2015).

\item Y. Mehmani, T. Sun, M.T. Balho, P. Eichhubl, and S. Bryant, Multiblock pore-scale modeling and 
upscaling of reactive transport: Application to carbon sequestration, Transp. Porous Media {\bf 95}, 
305 (2012).

\item J.P. Nogues, J.P. Fitts, M.A. Celia, and C.A. Peters, Permeability evolution due to dissolution 
and precipitation of carbonates using reactive transport modeling in pore networks, Water Resour. Res. 
{\bf 49}, 6006 (2013).

\item L. Li, C.A. Peters, and M.A. Celia, Upscaling geochemical reaction rates using pore-scale network 
modeling, Adv. Water Resour. {\bf 29}, 1351 (2006).

\item L.A. Dillard and M.J. Blunt, Development of a pore network simulation model to study nonaqueous 
phase liquid dissolution, Water Resour. Res. {\bf 36}, 439 (2000).

\item M. Hekmatzadeh, M. Dadvar, and M. Sahimi, Pore-network simulation of unstable miscible
displacements in porous media, Transp. Porous Media {\bf 113}, 511 (2016).

\item S. Ghanbarzadeh, M. Prodanovic, and M. A. Hesse, Percolation and grain boundary wetting in 
anisotropic texturally equilibrated pore networks, Phys. Rev. Lett. {\bf 113}, 048001 (2014).

\item P. Tahmasebi, F. Javadpour, and M. Sahimi, Multiscale and multiresolution modeling of shales
and their flow and morphological properties, Scientific Reports {\bf 5}, 16373 (2015).

\item P. Tahmasebi, M. Sahimi, A.H. Kohanpur, and A. Valocchi, Pore-scale simulation of flow of CO$_2$ 
and brine in reconstructed and actual 3D rock cores, J. Pet. Sci. Eng. {\bf 155}, 21 (2017).

\item B. Derrida and J. Vannimenus, A transfer-matrix approach to random resistor networks, J. Phys. A 
{\bf 15}, L557 (1982).

\item J. G. Zabolitzky, D. J. Bergman, and D. Stauffer, Precision calculation of elasticity for 
percolation, J. Stat. Phys. {\bf 44}, 211 (1986).

\item J.-M. Normand and H.J. Herrmann, Precise numerical determination of the superconducting exponent
of percolation in three dimensions, Int. J. Mod. Phys. C {\bf 1}, 207 (1990).

\item A.R. Mehrabi and M. Sahimi, Coarsening of heterogeneous media: application of wavelets, Phys. Rev. 
Lett. {\bf 79}, 4385 (1997)

\item F. Ebrahimi and M. Sahimi, Multiresolution wavelet coarsening and analysis of transport in
heterogeneous media, Physica A {\bf 316}, 160 (2002).

\item M. Sahimi, M. Naderian, and F. Ebrahimi, Efficient simulation of ac conduction in heterogeneous
materials at low temperatures, Phys. Rev. B {\bf 71}, 094208 (2005).

\item E. Pazhoohesh, H. Hamzehpour, and M. Sahimi, Numerical simulation of ac conduction in 
three-dimensional heterogeneous materials, Phys. Rev. B {\bf 73}, 174206 (2006).

\item A. Aghaei and M. Piri, Direct pore-to-core up-scaling of displacement processes: Dynamic pore 
network modeling and experimentation, J. Hydrol. {\bf 522}, 488 (2015).

\item J.K. Jasti, G. Jesion, and L. Feldkamp,  Microscopic imaging of porous media with x-ray computed 
tomography, SPE Form. Eval. {\bf 8}, 189 (1993).

\item M. A. Knackstedt, A. P. Sheppard, and M. Sahimi, Pore network modeling of two-phase flow in porous 
rock: The effect of correlated heterogeneity, Adv. Water Resour. {\bf 24}, 257 (2001).

\item T. T. Tsotsis, H. Patel, B. F. Najafi, D. Racherla, M. A. Knackstedt, and M. Sahimi, An overview 
of laboratory and modeling studies of carbon dioxide sequestration in coalbeds, Ind. Eng. Chem. Res. 
{\bf 43}, 2887 (2004).

\item Nvidia Corporation, NVIDIA CUDA programming Guide, 
http://docs.nvidia.com/cuda/cuda-c-programming-guide/abstract

\item G. R. Markall, D. A. Ham, and P. H. J. Kelly, Generating optimised finite element solvers for GPU 
architectures, AIP Conference Proceedings {\bf 1281}, 787 (2010).

\item P. Tahmasebi, M. Sahimi, G. Mariethoz, and A. Hezarkhani, Accelerating geostatistical simulations
using graphics processing units (GPU), Comput. Geosci. {\bf 46}, 51 (2012).

\item G.R. Markall, A. Slemmer, D.A. Ham, P. H. J. Kelly, C. D. Cantwell, and S. J. Sherwin, Finite 
element assembly strategies on multi- and many-core architectures, Int. J. Numer. Meth. Fluids {\bf 71}, 
8097 (2013). 

\item W. Liu, B. Schmidt, G. Voss, and W. Muller-Wittig, Accelerating molecular dynamics simulations 
using graphics processing units with CUDA, Comput. Phys. Commun. {\bf 179}, 634 (2008).

\item I. Ufimtsev and T. J. Martinez, Quantum chemistry on graphical processing units. 3. Analytical 
energy gradients, geometry optimization, and first principles molecular dynamics, J. Chem. Theor. Comput.
{\bf 5}, 2619 (2009).

\item R. Olivares-Amaya, M. A. Watson, R. G. Edgar, L. Vogt, Y. Shao, and A. Aspuru-Guzik, Accelerating 
correlated quantum chemistry calculations using graphical processing units and a mixed precision matrix 
multiplication library, J. Chem. Theor. Comput. {\bf 6}, 135 (2010).

\item M. Zheng, X. Li, and L. Guo,  Algorithms of GPU-enabled reactive force field (ReaxFF) molecular 
dynamics, J. Mol. Graphics Model. {\bf 41}, 1 (2013).

\item  F.J. Molz, H.H. Liu, and J. Szulga, Fractional Brownian motion and fractional Gaussian noise in
subsurface hydrology: a review, presentation of fundamental properties, and extensions, Water Resour. 
Res. {\bf 33}, 2273 (1997).

\item M. Sahimi and S. Mukhopadhyay, Scaling properties of a percolation model with long-range
correlations, Phys. Rev. E {\bf 54}, 3870 (1996).

\item M. Sahimi, Long-range correlated percolation and flow and transport in heterogeneous porous media,
J. de Physique I {\bf 4}, 1263 (1994).

\item M.A. Knackstedt, A.P. Sheppard, and W.V. Pinczewski, Simulation of mercury porosimetry on
correlated grids: evidence for extended correlated heterogeneity at the pore scale in rocks, Phys. Rev. 
E {\bf 58}, R6923 (1998). 

\item H. Dashtian, G.R. Jafari, M. Sahimi, and M. Masihi, Scaling, multifractality, and long-range 
correlations in well log data of large-scale porous media, Physica A {\bf 390}, 2096 (2011).

\item H. Dashtian, Y. Yang, and M. Sahimi, Nonuniversality of the Archie exponent due to multifractality 
of resistivity well logs, Geophys. Res. Lett. {\bf 42}, 10655 (2015).

\item B.B. Mandelbrot and J.W. van Ness, Fractional Brownian motion, fracrional Guassian noise, and their
applications, SIAM Rev. {\bf 10}, 422 (1968).

\item M. Ansari-Rad, S.M. Vaez Allaei, and M. Sahimi, Nonuniversality of roughness exponent of
quasi-static fracture surfaces, Phys. Rev. E {\bf 85}, 021121 (2012). 

\item N. Wilt, {\it The CUDA Handbook: a Comprehensive Guide to GPU Programming} (Addison-Wesley, New 
York, 2013).

\item http://www.nvidia.com/object/nvidia-kepler.html

\item R. Barrett, M. Berry, T.F. Chan, J. Demmel, J. Donato, J. Dongarra, V. Eijkhout, R. Pozo, C.
Romine, and H. van der Vorst, {\it Templates for the Solution of Linear Systems: Building Blocks for 
Iterative Methods}, 2nd ed. (SIAM, Philadelphia, 1994).

\item S. Kirkpatrick, Percolation and conduction, Rev. Mod. Phys. {\bf 45}, 574 (1973).

\end{list}%

\newpage

\begin{figure}[t]
\begin{center}
\includegraphics[scale=0.7]{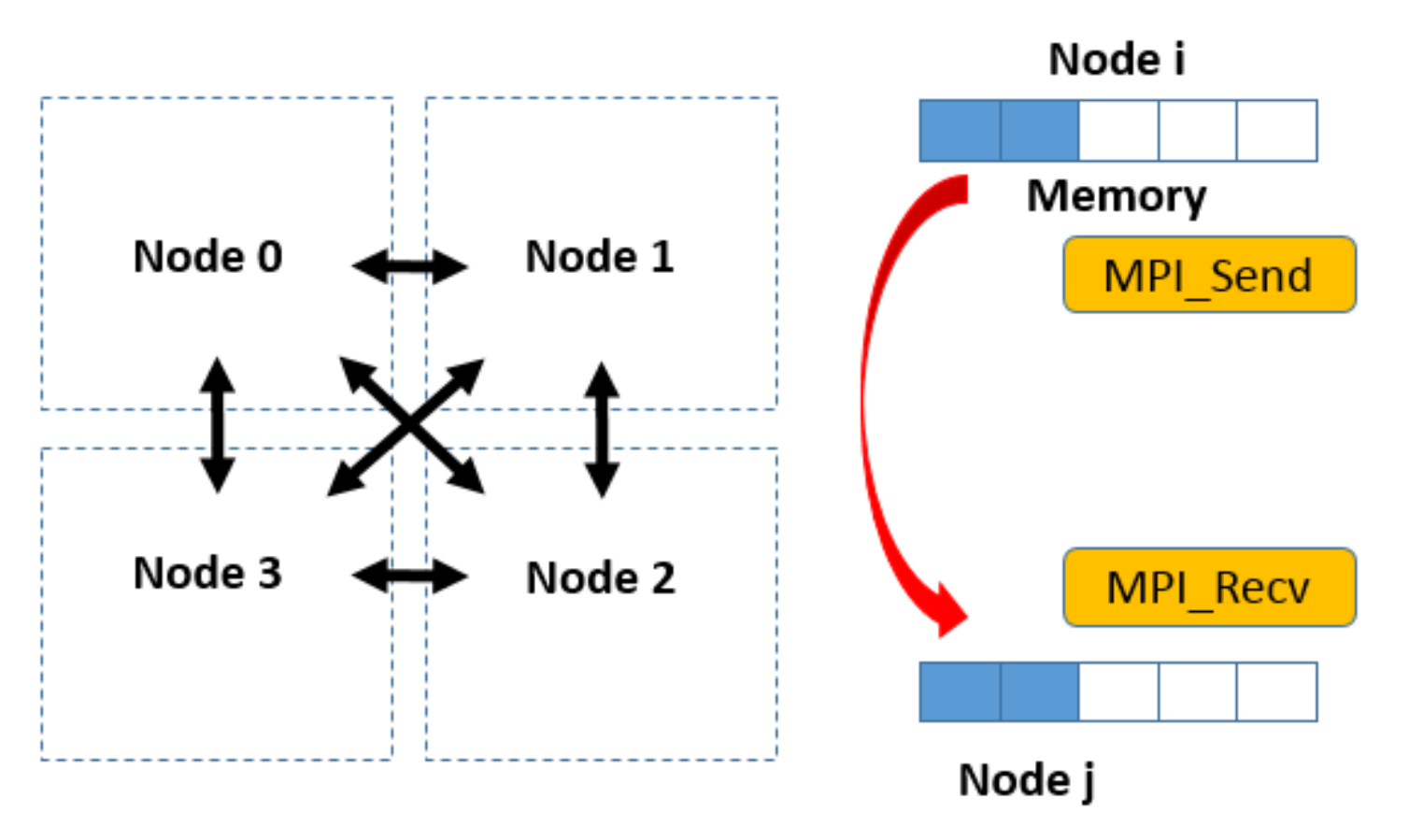}
\caption{Communications between four nodes (left) and point-to-point communication model.}
\end{center}
\end{figure}

\newpage

\begin{figure}[t]
\begin{center}
\includegraphics[scale=0.6]{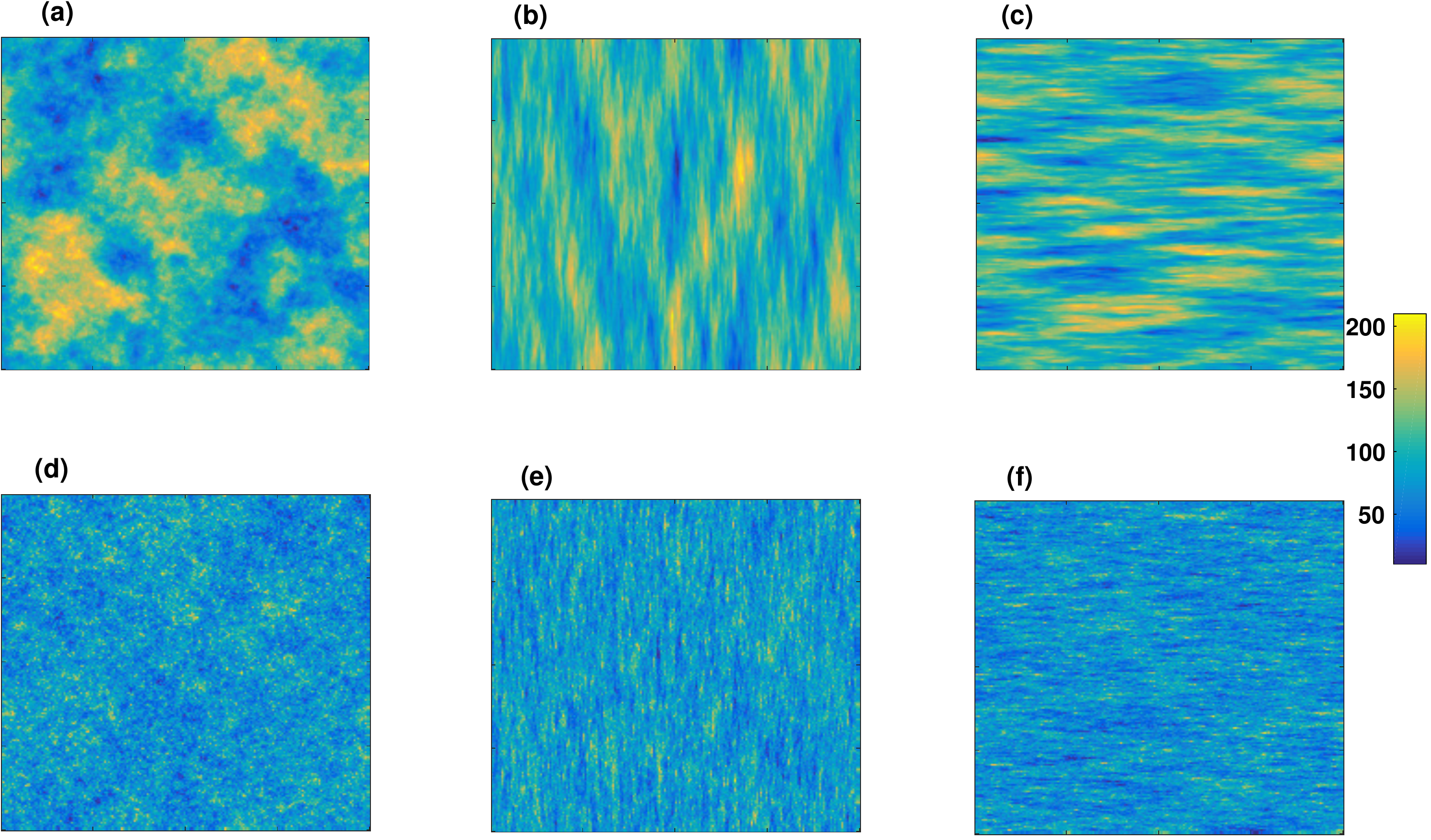}
\caption{Examples of the networks with correlated conductances. The top and bottom rows show, 
respectively, the networks with the Hurst exponent of $H=0.75$ and 0.35, and $\beta_x/\beta_y$ of 1 [(a) 
and (d)]; 1/5 [(b) and (e)], and 5 [(c) and (f)].} 
\end{center}
\end{figure}

\newpage

\begin{figure}[t]
\begin{center}
\includegraphics[scale=0.7]{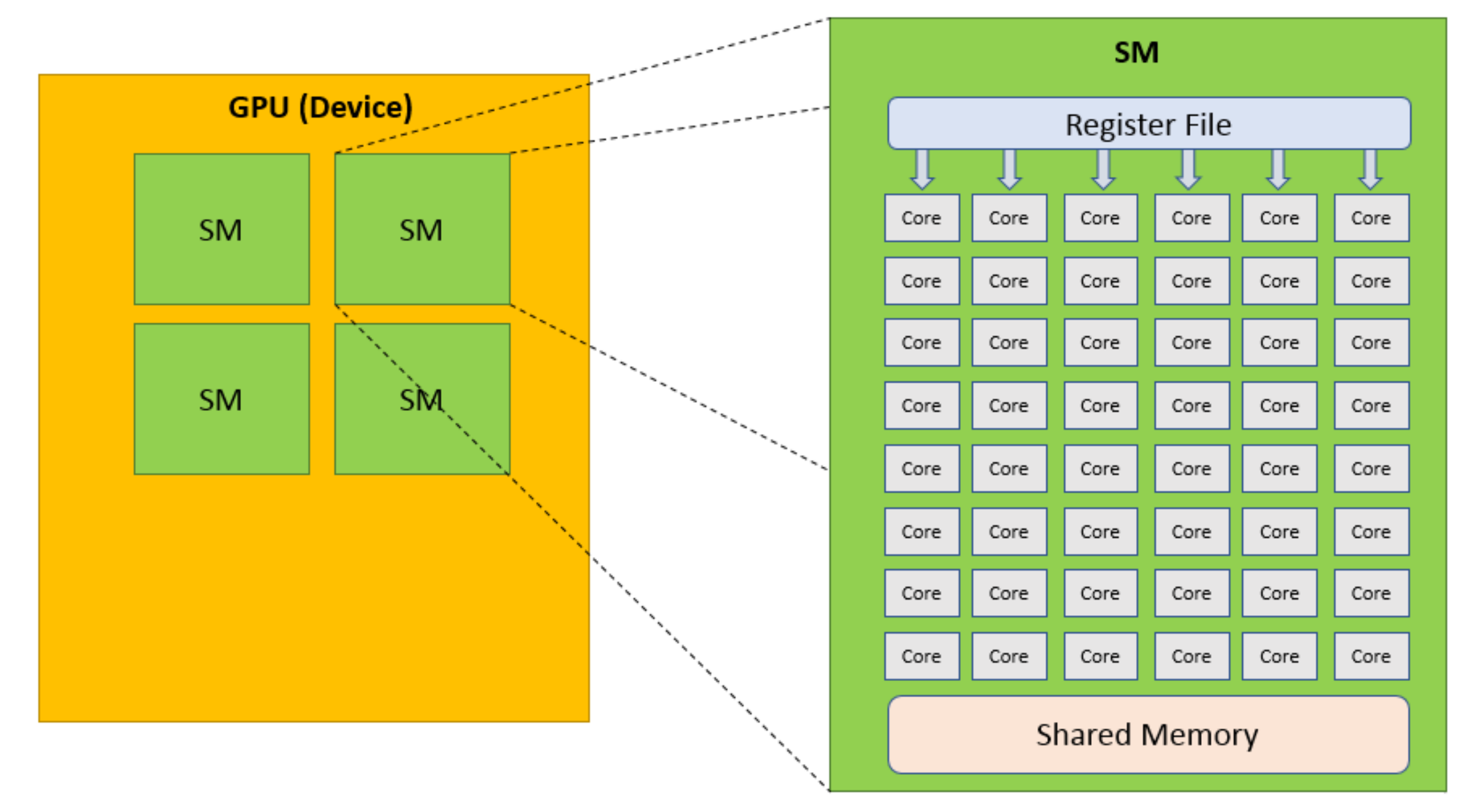}
\caption{Hierarchy of GPU architecture with four streaming multiprocessors (SM).}
\end{center}
\end{figure}

\newpage

\begin{figure}[t]
\begin{center}
\includegraphics[scale=0.7]{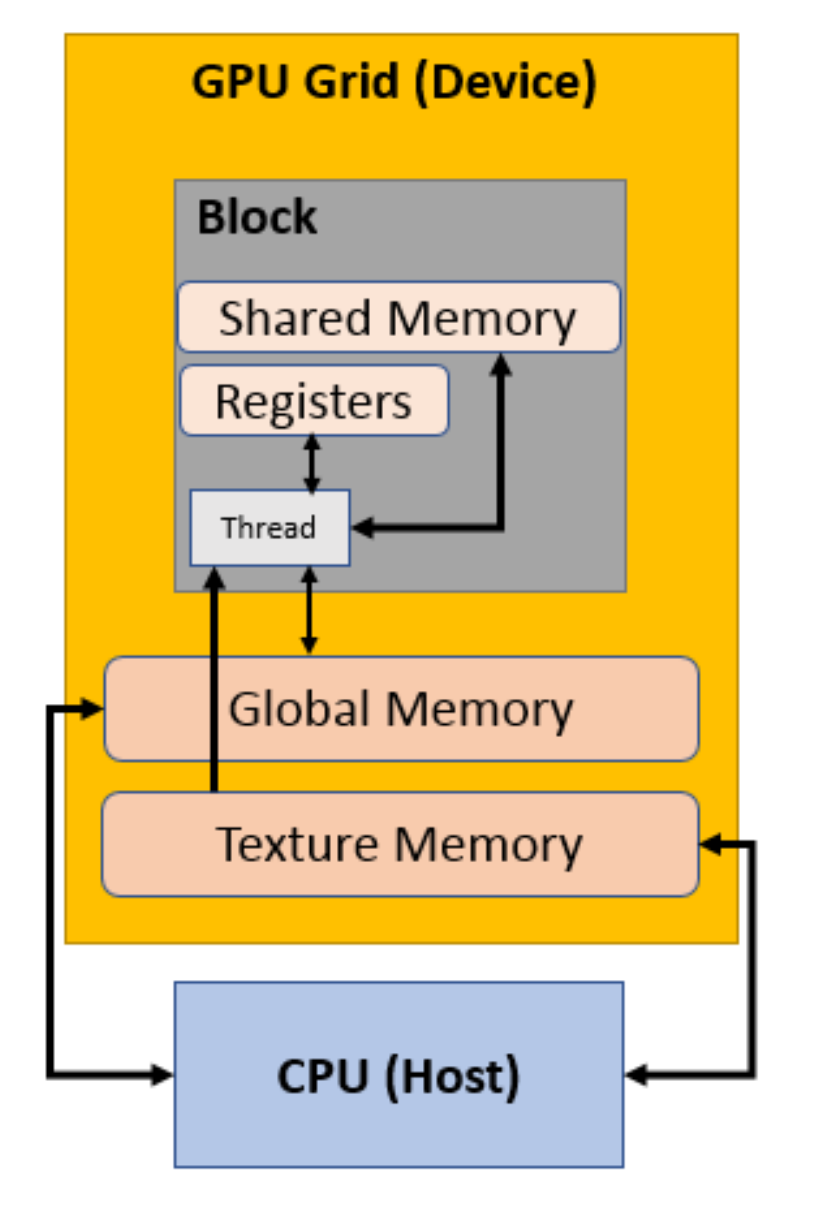}
\caption{The CUDA memory model.}
\end{center}
\end{figure}

\newpage

\begin{figure}[t]
\begin{center}
\includegraphics[scale=0.6]{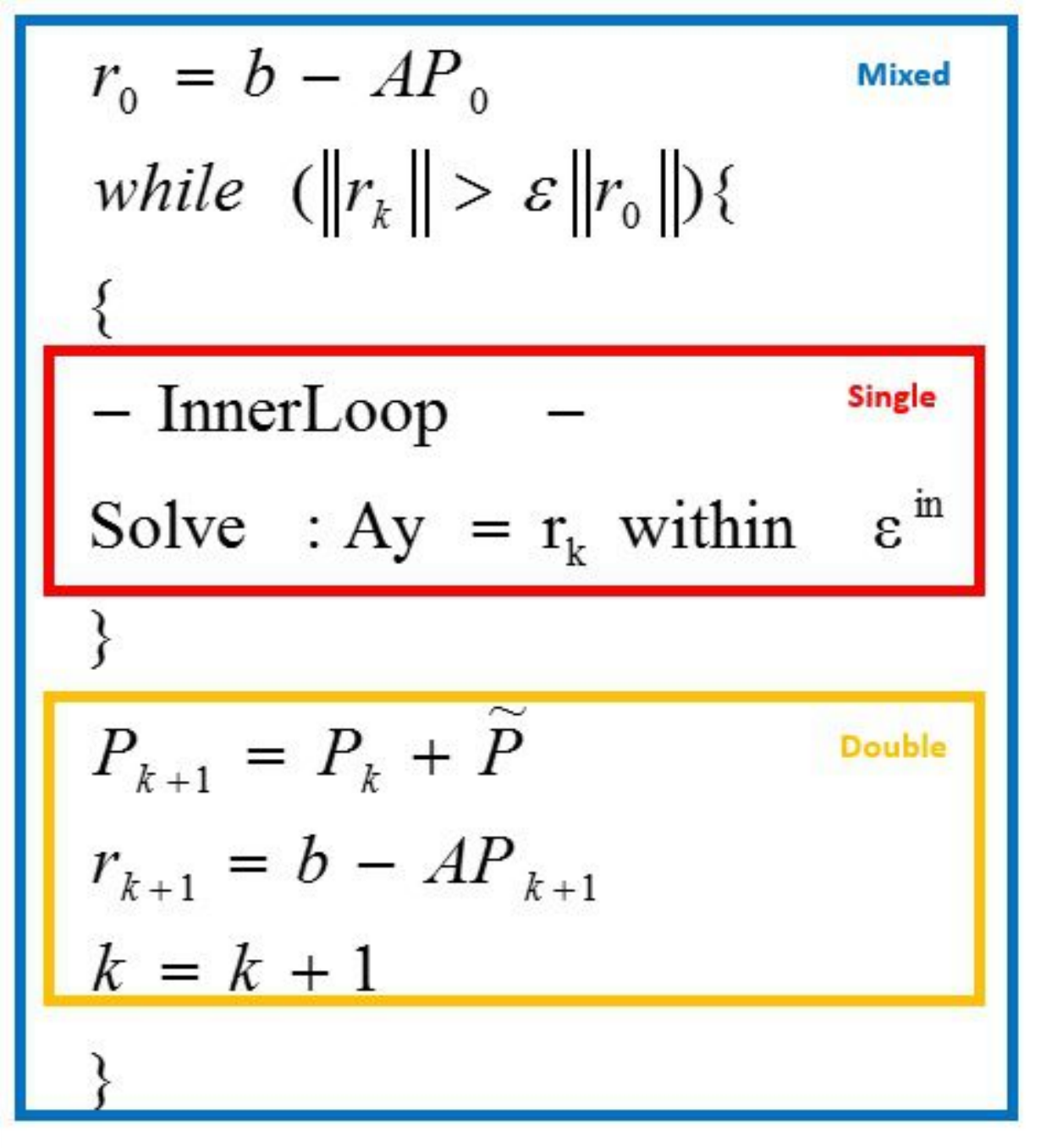}
\caption{Mixed-precision algorithm for the conjugate-gradient method that was used in the computations.}
\end{center}
\end{figure}

\newpage

\begin{figure}[t]
\begin{center}
\includegraphics[scale=0.6]{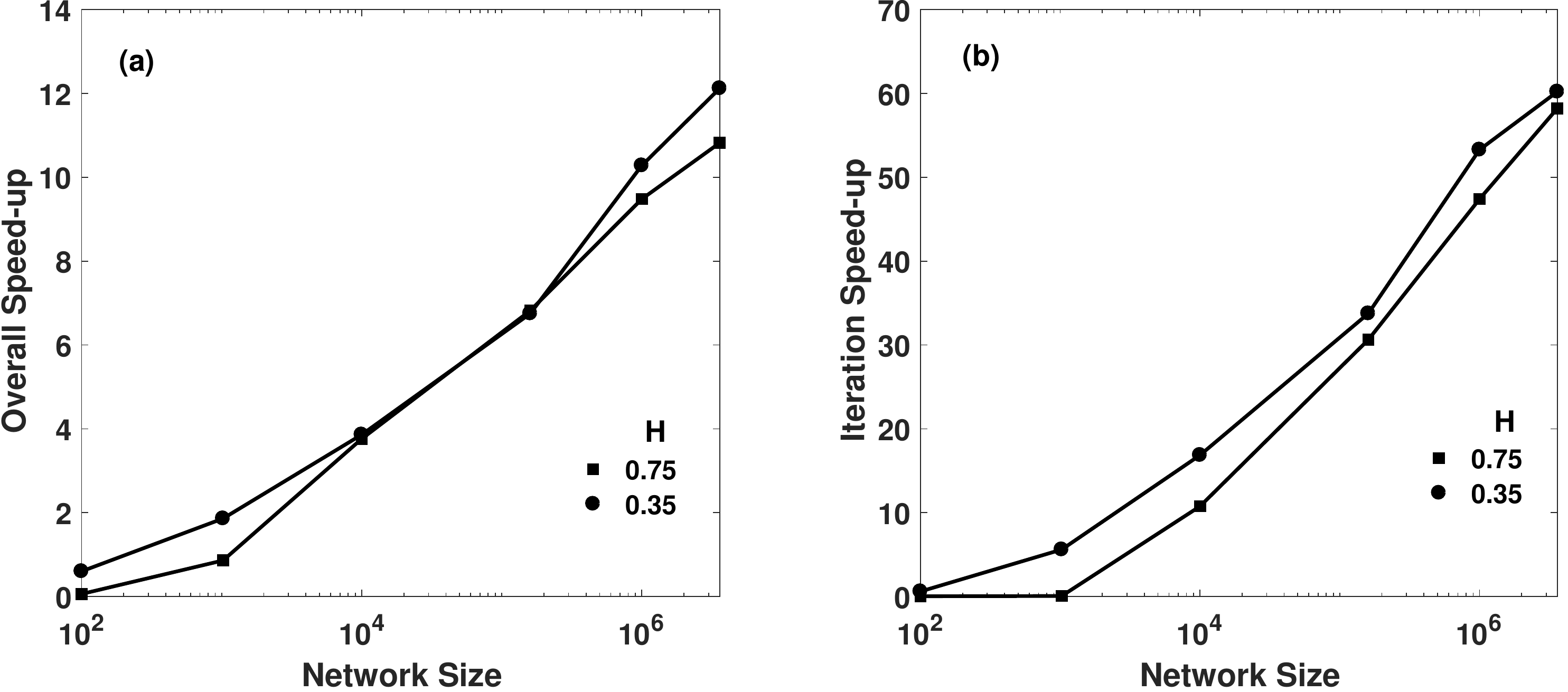}
\caption{(a) Dependence of the overall speed-up on the size of isotropic networks and the Hurst exponent 
$H$. (b) Speed up for one iteration in CG algorithm.}
\end{center}
\end{figure}

\newpage

\begin{figure}[t]
\begin{center}
\includegraphics[scale=0.6]{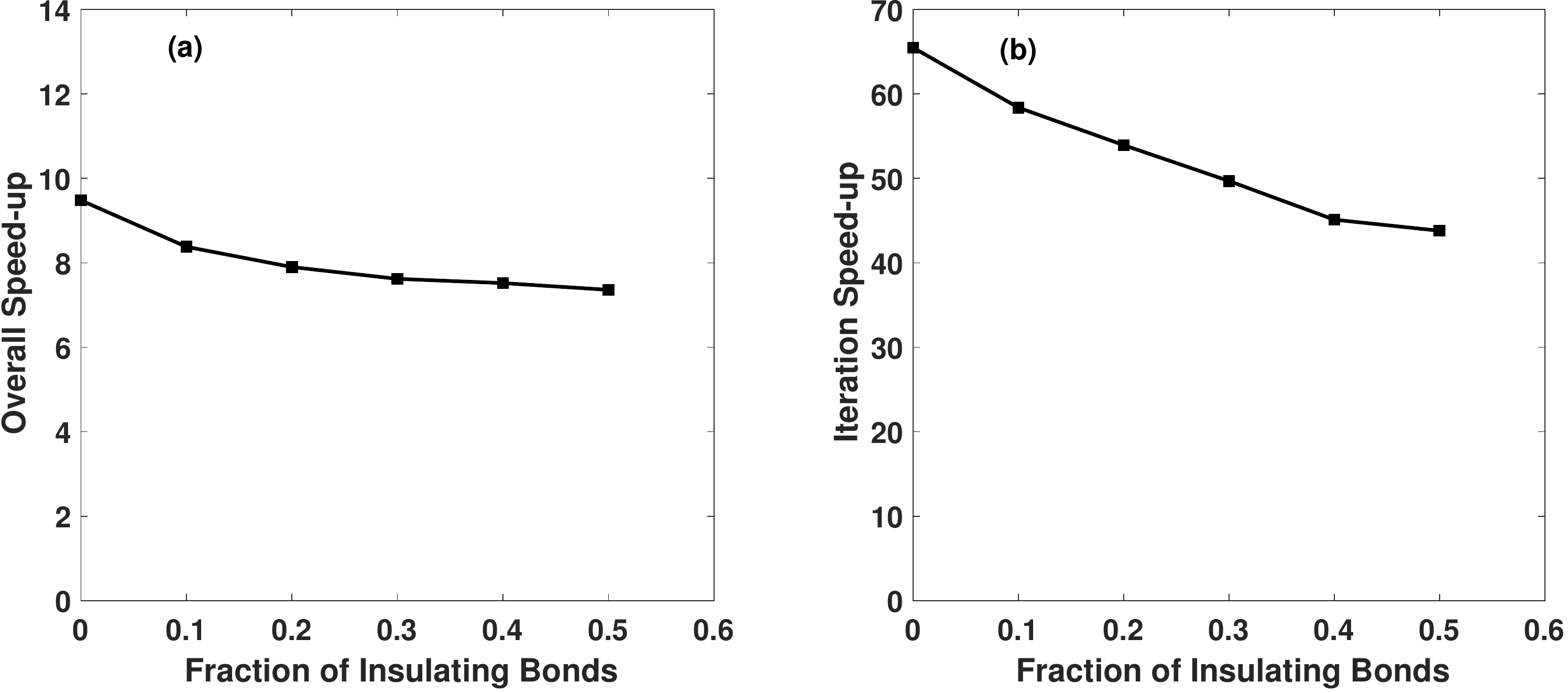}
\caption{(a) The overall speed-up of the computations with a random resistor network with $3.6\times 
10^6$ nodes. (b) Speed up for one iteration in CG algorithm.}
\end{center}
\end{figure}

\newpage

\begin{figure}[t]
\begin{center}
\includegraphics[scale=0.6]{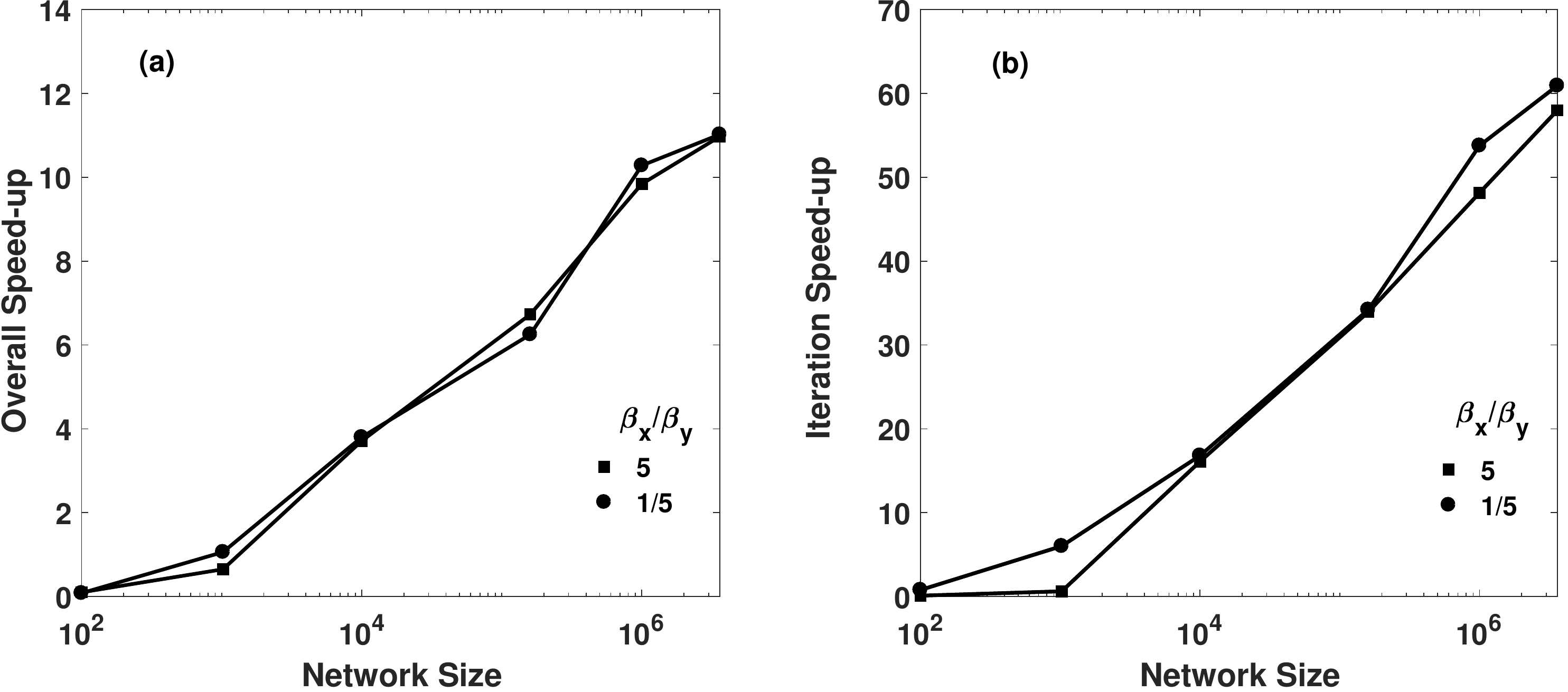}
\caption{(a) Dependence of the overall speed-up on the size of anisotropic networks and the anisotropy 
parameter $\beta_x/\beta_y$. The Hurst exponent is $H=0.35$. (b) Speed up for one iteration in CG
algorithm.}
\end{center}
\end{figure}

\newpage

\begin{figure}[t]
\begin{center}
\includegraphics[scale=0.6]{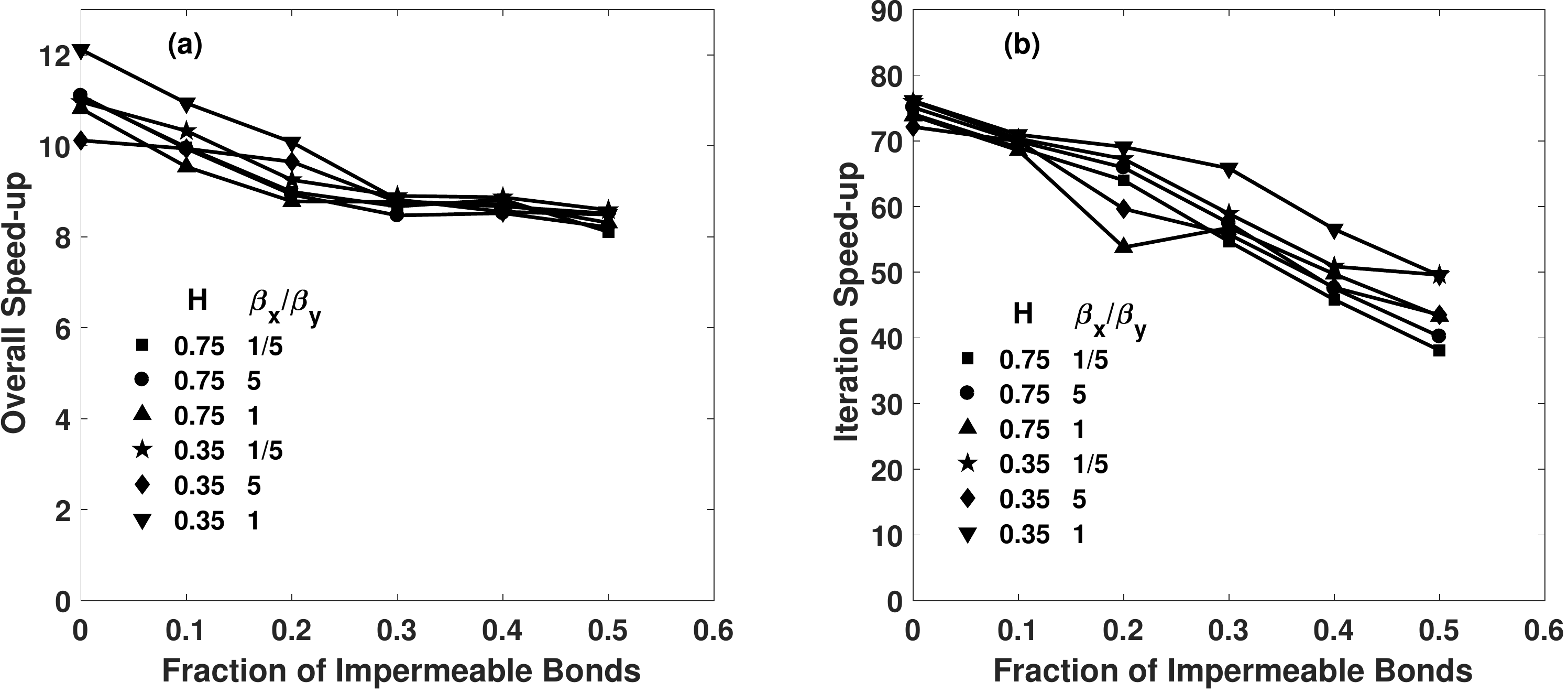}
\caption{a) Dependence of the overall speed-up of the computations for pore networks with $3.6 \times
10^6$ nodes on the fraction of the permeable bonds, the Hurst exponent $H$ and the anisotropy parameter 
$\beta_x/\beta_y$. (b) Speed up for one iteration in CG algorithm.}
\end{center}
\end{figure}

\newpage

\begin{figure}[t]
\begin{center}
\includegraphics[scale=0.6]{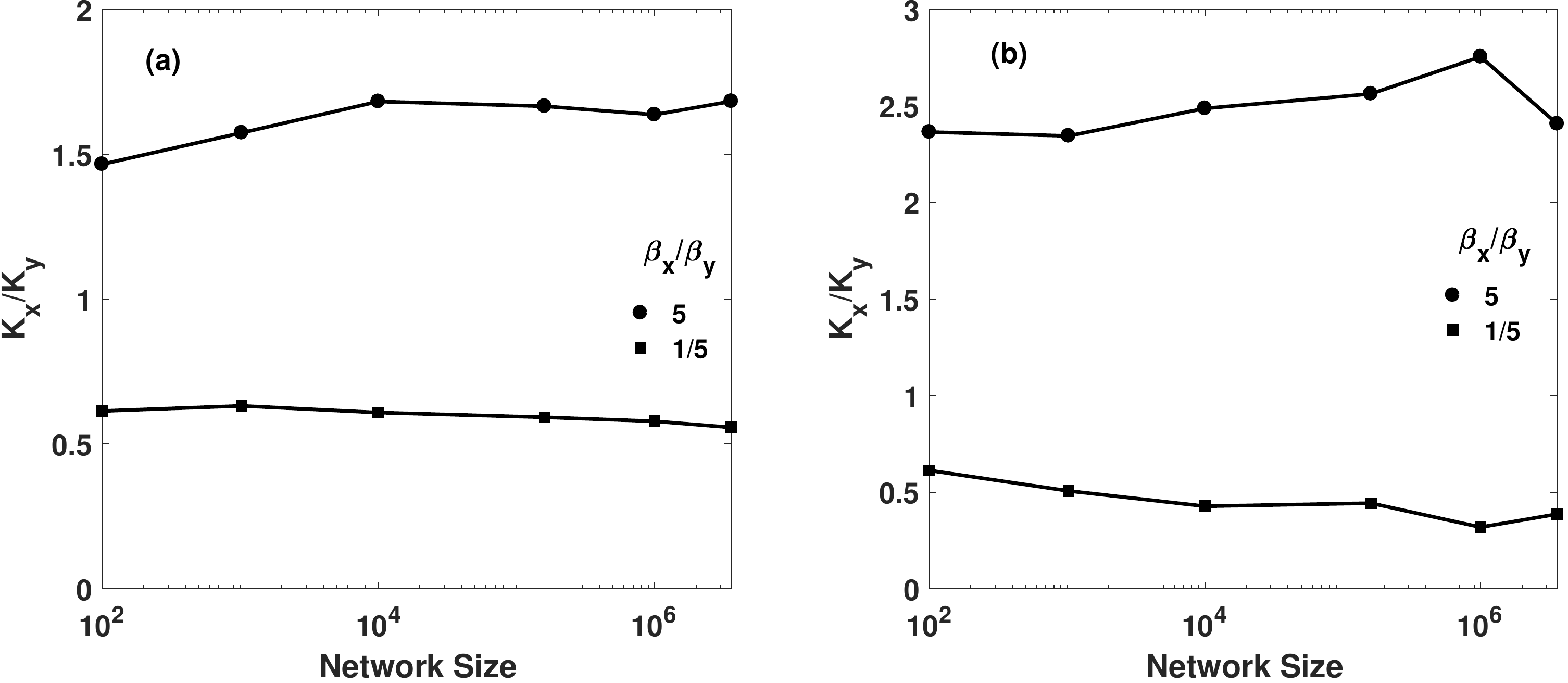}
\caption{Approximate bounds for the permeability ratio $K_x/K_y$ and their dependence on the anisotropy 
ratio $\beta_x/\beta_y$ and the Hurst exponent $H$ for (a) $H=0.75$ and (b) $H=0.35$.}
\end{center}
\end{figure}

\end{document}